\documentclass{emulateapj}

\newcommand{\simle}{\mbox{$\stackrel{<}{_{\sim}}$}}

\newcommand\msun{\hbox{\,M$_\odot$}}

\shorttitle{RS Oph Interferometry}
\shortauthors{Monnier et al.}

\begin{document}


\title{No Expanding Fireball: Resolving the Recurrent Nova RS~Ophiuchi with Infrared Interferometry}


\author{J.~D.~Monnier\altaffilmark{1}, 
R.~K.~Barry\altaffilmark{2,12}, 
W.~A.~Traub\altaffilmark{7,8},
B.~F.~Lane\altaffilmark{3}, 
R.~L.~Akeson\altaffilmark{6},
S.~Ragland\altaffilmark{4},  
P.~A.~Schuller\altaffilmark{7}, 
H.~Le~Coroller\altaffilmark{11},
J.-P.~Berger\altaffilmark{5},
R.~Millan-Gabet\altaffilmark{6},
E.~Pedretti\altaffilmark{1},
F.~P.~Schloerb\altaffilmark{9},
C.~Koresko\altaffilmark{6},
N.~P.~Carleton\altaffilmark{7},
M.~G.~Lacasse\altaffilmark{7}, 
P.~Kern\altaffilmark{5}, 
F.~Malbet\altaffilmark{5},
K.~Perraut\altaffilmark{5},
M.~J.~Kuchner\altaffilmark{12}
and
M.~W.~Muterspaugh\altaffilmark{10}
}

\altaffiltext{1}{monnier@umich.edu: University of Michigan Astronomy Department, 941 Dennison Bldg, Ann Arbor, MI 48109-1090, USA.}
\altaffiltext{2}{Johns-Hopkins University}
\altaffiltext{3}{Massachusetts Institute of Technology}
\altaffiltext{4}{W. M. Keck Observatory}
\altaffiltext{5}{Laboratoire d'Astrophysique de Grenoble}
\altaffiltext{6}{Michelson Science Center (Caltech)}
\altaffiltext{7}{Harvard-Smithsonian Center for Astrophysics}
\altaffiltext{8}{Jet Propulsion Laboratory (Caltech)}
\altaffiltext{9}{University of Massachusetts, Amherst}
\altaffiltext{10}{California Institute of Technology}
\altaffiltext{11}{Observatoire de Haute-Provence, France}
\altaffiltext{12}{NASA -- Goddard Space Flight Center}

\begin{abstract}
  Following the recent outburst of the recurrent nova RS~Oph
  on 2006 Feb 12,
  we measured its near-infrared size using the IOTA, Keck, and
  PTI Interferometers at multiple epochs.  The characteristic size of
  $\sim$3~milliarcseconds hardly changed over the first 60 days of the
  outburst, ruling out currently-popular models whereby the
  near-infrared emission arises from hot gas in the expanding shock.
  The emission was also found to be significantly asymmetric,
  evidenced by non-zero closure phases detected by IOTA.  The physical
  interpretation of these data depend strongly on the adopted distance
  to RS~Oph.  Our data can be interpreted as the first direct
  detection of the underlying RS~Oph binary, lending support to the
  recent ``reborn red giant'' models of Hachisu \& Kato.  However,
  this result hinges on an RS~Oph distance of $\simle$540pc, in strong
  disagreement with the widely-adopted distance of $\sim$1.6~kpc.  At
  the farther distance, our observations imply instead the existence of
  a non-expanding, dense and ionized circumbinary gaseous disk 
  or reservoir responsible for the bulk of the near-infrared emission.  
  Longer-baseline infrared interferometry is uniquely
  suited to distinguish between these models and to ultimately
  determine the distance, binary orbit, and component masses for
  RS~Oph, one of the closest-known (candidate) SNIa progenitor systems.

\end{abstract}




\section{Introduction}

Most astronomers are familiar with Classical Novae, exploding stars in
which an accreting white dwarf (WD) in an interacting binary system
accumulates enough material for it to become unstable to hydrogen
burning.  The expanding blastwave from one such event Nova Aql 2005
was recently observed by infrared (IR) interferometry
\citep{lane2005aas} and the geometric distance was estimated based on
velocities from spectral line observations.  This result is consistent
with the ``optically-thick fireball'' model which has been
successfully used for twenty years to explain the time-evolution of
the spectral energy distribution of classical novae \citep{gehrz1988}.

While classical novae are expected to recur, very few actually have in
recorded history.  RS~Oph is one of the handful of so-called
``recurrent novae'' with (now) 6~outbursts since 1898
\citep{warner1976}. The most recent outburst occurred on February 12,
2006 \citep{narumi2006}, and this unusual event motivated intense
monitoring  by the IR interferometry community.

The special nature of RS~Oph is thought to stem from two
causes. First, the WD is likely extremely close to
the Chandrasekhar limit, since the amount of hydrogen needed to trigger
an outburst decreases dramatically as the WD mass increases.  Indeed,
detailed models indicate the WD mass is within 1\% of exploding as a
type Ia supernova \citep[e.g.,][]{hachisu2001}. 
Second, the mass-losing companion for RS~Oph is a red giant (RG) with a wind,
providing a high density medium for accretion onto the WD as well as for 
the exploding blastwave to
interact with.  \citet{bode1985} 
have produced the most successful model
for recurrent novae, drawing a clear analogy to extragalactic
supernovae and explaining the radio and X-ray light curves in this
context.

\citet{evans1988}
were the first to study in detail the IR time
evolution of a recurrent nova.  They monitored closely the 1-3.5
micron flux of RS~Oph for about 3 years after the 1985 eruption.  They
found the light curve had a characteristic (2-mag) decay time scale of about
30~days, and compared their observations to the generic predictions of
the \citet{bode1985} model.  They concluded that their observations
could come from the hot, post-shock gas -- this model would predict 
that the IR source should be seen linearly expanding
at a rate of about $\sim$1~milliarcseconds per day at a distance of 1.6kpc.
This distance estimate is based (most securely) on the 
expanding size of the radio emission observed in
1985 by \citet{hjellming1986} and \citet{taylor1989}, 
assuming 
association with the forward shock \citep[new radio data 
re-confirm the 1985 observations;][]{rupen2006,obrien2006}.
As will become clear, the distance to RS~Oph is key to the
interpretation of the IR interferometry data presented here.

Challenging this interpretation, \citet[][see also Kato
1991]{hachisu2001} have recently produced a comprehensive model for
recurrent novae meant to explain a wide range of the known novae
properties, and makes a specific prediction for the origin of the
near-infrared (NIR) continuum which is very different from
\citet{evans1988}.  Following onset of the thermonuclear runaway of
the hydrogen shell around the WD, the shell expands to AU~size, in
effect turning the WD back into a red giant.  The shell stably burns
hydrogen for a few weeks, shrinking back to
the size of white dwarf. According to this model, hot post-shock gas
plays no role in forming the IR continuum.  Furthermore,  \citet{hachisu2001} 
prefer a much closer distance of 600~pc
implying a binary separation of 2.9 milliarcseconds, easily detectable
with current interferometers.

In this Letter, we report first-ever size measurements for RS~Oph in
the NIR using long-baseline interferometry. Our results are
surprising, ruling out the favored expanding fireball model, raising
doubts about the established distance to RS~Oph, and motivating a new
model for the NIR emission.

\section{Observations}
\label{section:observations}
In the Letter we report on data from 3 different interferometers and 
a summary of observations can be found in 
Table~\ref{table_obslog}.  Here we briefly introduce each dataset.

Most of our data were obtained at the Infrared-Optical Telescope Array
\citep[IOTA,][]{traub2003} which has baselines between 5--38m.
The IONIC3 combiner \citep{berger2003} was used to measure
3~visibilities (V$^2$) and 1~closure
phase (CP) simultaneously in the broadband H band filter ($\lambda_0 =
1.65\mu$m, $\Delta\lambda = 0.3\mu$m). Data analysis
procedures have been documented in recent papers
\citep{monnier2004b,monnier2006}.  For the data here, 
we have adopted a calibration error
$\Delta$V$^2=$ 0.05 (relative error) on baselines AB and AC and
$\Delta$V$^2=$ 0.10 for baseline BC.

The Keck Interferometer (KI) was used to observe RS~Oph on a single
night 4 days after the burst (UT 2006 Feb 16) with a baseline of
$\sim$85m.  Facility and instrument descriptions can be found in
recent KI publications \citep{colavita2003,cw2003,cw2000}. 
The K-band ($\lambda_0 = 2.18\mu$m, $\Delta\lambda = 0.3\mu$m) data
reported here were the by-product of a nulling observation (being
prepared for a separate publication) and the calibration sequence is
somewhat modified from the standard procedures; in particular, only
one ratio measurement was made per integration.

Lastly, the Palomar Testbed Interferometer (PTI) observed RS~Oph on
3~nights in 2006 April using the K-band detection system ($\lambda_0 = 2.20\mu$m, $\Delta\lambda = 0.4\mu$m) with a $\sim$85~m baseline (oriented in NE direction).
Detailed instrument and data analysis descriptions
for PTI can be found in the literature  \citep{colavita1999,colavita99}.  
Because of the inherent
faintness of the source in April, coherent integration was used for analysis
and a large calibration error of $\Delta$V$^2$=0.10 (absolute, not
relative) was added in quadrature with the internal error for model
fitting in this paper.

We have split the data into 3 epochs -- from 2006 Feb 16 to 23 (Days
4-11), from Feb 26 to Mar 13 (Days 14-29), and from April 2 to April
18 (Days 49-65).  The UV-averaged visibility data for each epoch are
presented in Figure~\ref{fig1} along with some Gaussian profiles for
comparison.  The IOTA closure phase results are shown in
Figure~\ref{fig2}, also split into the three epochs.

All V$^2$ and closure phase data are available 
from the authors;
all data products are stored in the FITS-based, optical interferometry
data exchange format (OI-FITS), recently described in 
\citet{pauls2005}.

\section{Analysis}

\begin{figure}[hbt]
\begin{center}
\includegraphics[angle=90,width=6in]{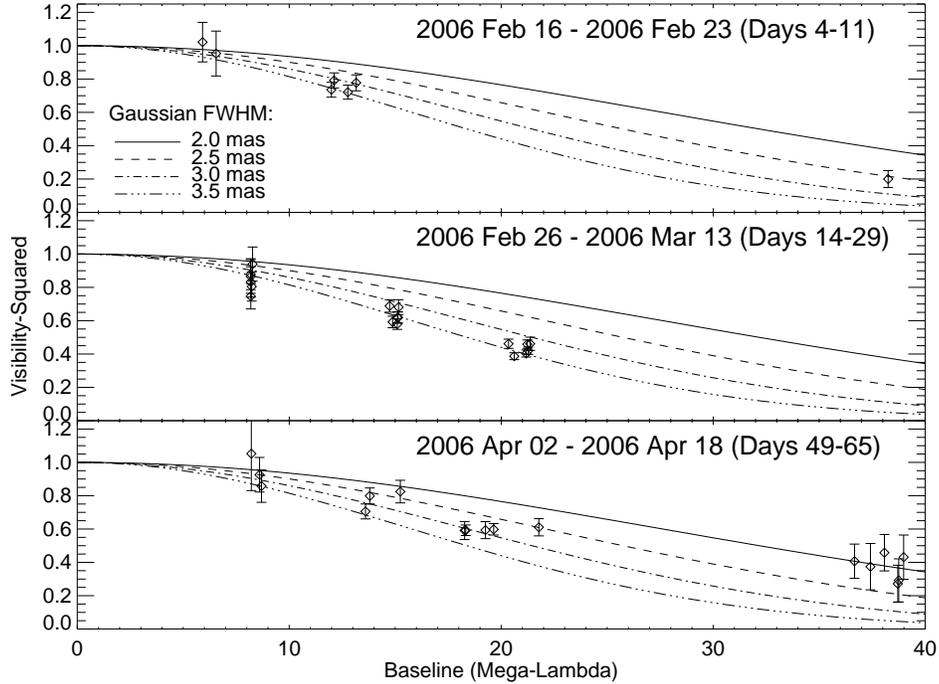}
\figcaption{\footnotesize 
This figure shows UV-averaged visibility data for RS~Oph split into three different time periods.
The x-axis shows the spatial frequency (projected baseline in units of wavelength)  
while the y-axis shows the visibility-squared.   Four curves representing Gaussian profiles are
also included to show the characteristic size and to allow intercomparison of data in the different panels.
For reference, all data shortward of 25M$\lambda$ derive from IOTA, while
longer baseline data come from KI (Epoch I) and PTI (Epoch III).
\label{fig1}}
\end{center}
\end{figure}

\begin{figure}[hbt]
\begin{center}
\includegraphics[angle=90,width=6in]{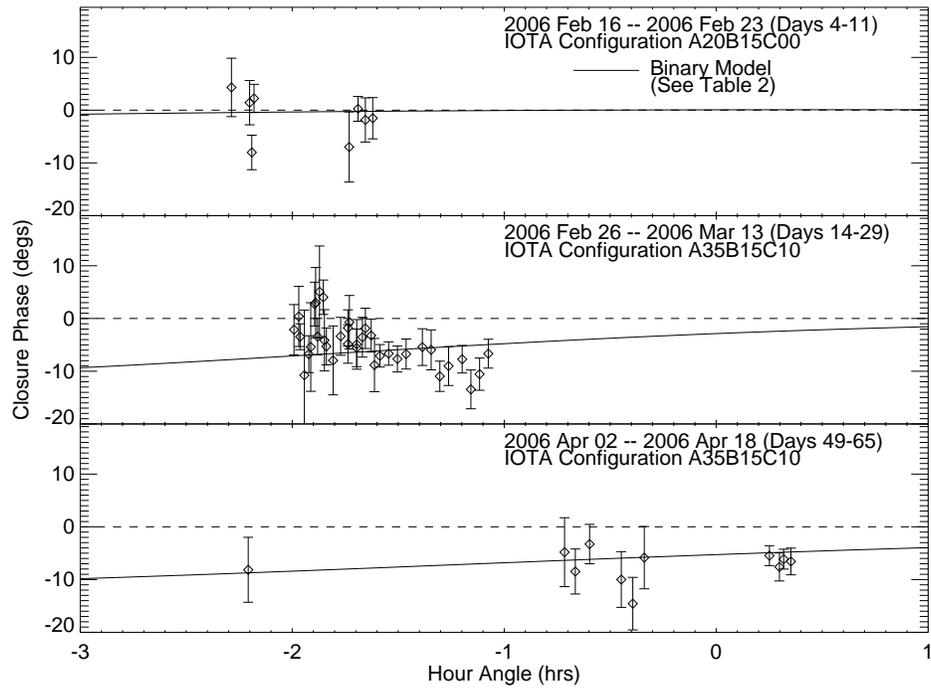}
\figcaption{\footnotesize 
This figure shows IOTA closure phase data for RS~Oph split into three different time periods.
The x-axis shows the hour angle of the observations 
while the y-axis shows the observed closure phase.   The solid line shows expected closure phase for
the binary model parameters found in Table~2 and discussed in \S\ref{binary}.  The
dashed line shows the expected closure phase signal for the symmetric Gaussian model.
\label{fig2}}
\end{center}
\end{figure}

Inspection of Figure~\ref{fig1} reveals that the
visibility curve for RS~Oph changes very little between day 4 and 65
(since the outburst).  This is surprising since
the H band brightness faded by a factor $>$10 during
this time.  This result is discordant
with the generic prediction of
\citet{evans1988} who interpreted the IR light curve in terms of
the time-evolution of post-shock gas at 10$^5$K moving at the speed of
the contact discontinuity, 1400~km/s \citep[following the earlier work
by][]{bode1985}. 
This model requires the IR emission to be seen expanding
at a rate of $\sim$1.0~milliarcseconds per day (assuming a distance
of 1.6kpc), an interpretation that is now ruled out.

Before discussing alternative models in \S\ref{discussion}, we wish to
carry-out some model fitting to the interferometry data.  Here we only
consider two simple models -- a circularly symmetric Gaussian and a
binary star model.  For all fits and calculation of reduced $\chi^2$,
we have used the original data points before UV-averaging.

\label{gauss}

First, we fit a circularly symmetric Gaussian to each epoch of data, split by
wavelength.  Table~\ref{table_model} contains the best-fitting
Full-Width at Half-Maxima (FWHM) and the reduced $\chi^2$ 
for both the V$^2$ and CP (the model CP is always
zero for a Gaussian profile).  The Gaussian
model is a reasonable fit for the IOTA visibility alone, but clearly
cannot fit the non-zero closure phase seen in March and April.  Also,
Figure~\ref{fig1} shows that no good fit was possible when combining
IOTA with Keck and PTI data, indicating this model is too simplistic
to explain the full range of baselines and/or the
wavelength-dependence. Sizes derived from the
longer-baseline K-band data are systematically smaller than those
derived from shorter-baseline H-band data (IOTA).

\label{binary}
As discussed in the Introduction, \citet{hachisu2001} suggest that the
nova's IR light curve might be due primarily to a rapid increase in
brightness of the WD as it returns, albeit briefly, to a red giant
phase.  Motivated by this work, we realized that the IR emission might
be due to the underlying RS~Oph binary itself, and this might explain
the general puzzling features of our data: non-expanding emission
size, the inadequacies of the Gaussian fit, and the non-zero closure
phase.  

In order to test this idea, we fit binary models to the data for each of the
three time periods, treating the
brightness ratio as independent of wavelength in order to fit the H and K
band data together.   The IOTA, KI, and PTI complement each other in Fourier
coverage and an exhaustive grid search of separations less than 10~mas 
found unique binary star
solutions \footnote{Our limited (u,v) coverage
admit some unlikely additional solutions with larger binary separations which will be discussed fully in a future modeling paper.}.
Table~\ref{table_model} contains the best-fit binary models for the
three time periods, including the reduced $\chi^2$ (V$^2$, CP).  All
three epochs are reasonably fit by a similar binary model.  The only
parameter that changed significantly between the epochs was the
brightness ratio.  The closure phase predictions for the binary models
are plotted along with the closure phase data in Figure~\ref{fig2}.
We note that our uv-plane is missing coverage in the NW direction and thus
elongated structure in this direction would be observed fore-shortened.

\section{Discussion}
\label{discussion}

Because the expanding fireball model fails to 
explain the nearly static size scale of the IR emission,
we now seek suitable alternative emission mechanisms for the
time-variable IR emission from the recurrent nova RS~Oph.  We have
pursued the reborn red giant (thermonuclear runaway) model of
\citet{hachisu2001} and found that indeed our 3-interferometer
combined dataset can be explained by a simple binary model with
separation of $\sim$3.2~milliarcsecond, PA ~30$\arcdeg$ E of N, and a
brightness ratio varying from 2.5:1 to 5:1.  Next we subject the
binary hypothesis to further scrutiny.

\subsection{Binary Interpretation of Near-Interferometry Data}
\label{binary2}
Based on single-line radial velocity data, \citet{fekel2000} finds the
RS~Oph binary orbit to be roughly circular with a period of
455.72$\pm$0.83 days and mass function $f=0.221\pm0.038$~\msun.  RGs in 
symbiotic systems are typically 1-3\msun
\citep{dobrzycka1994} and we expect recurrent novae to contain a
Chandrasekhar mass WD (1.4\msun); these facts combined with the known
mass function rule out RG masses greater than 2\msun.
Assuming the RS~Oph system mass to be 2.4-3.4$\msun$, we find the
component separation to be 1.55-1.74~AU, or (unprojected) 
2.59-2.90~mas at the 600pc
distance preferred by \citet{hachisu2001} -- only slightly
smaller than our observed separation of 3.2~mas.

Since the RS~Oph outburst took place only 1 month before maximum 
redshifted velocity \citep{fekel2000}, our measured binary parameters
represent the true orbital semi-major axis and orbital $\Omega$ for
RS~Oph with only weak $sin(i)$ dependencies. Thus, a small reduction
in the distance estimate (540~pc) brings the interferometer binary
model in agreement with expectations from Kepler's laws.

The binary model fits (Table~\ref{table_model}) show 
evidence for a change in the
brightness ratio over time.  While the \citet{hachisu2001} theory
predicts a time-changing brightness ratio, it is beyond the scope of this
paper to test the compatibility with the observed IR light curves due to 
complications from the role of the irradiated RG photosphere and the presence of
a post-outburst WD accretion disk.

\subsection{Circumbinary Reservoir of Hot Gas}
The distance estimate of $\simle$540~pc derived in the last section
stands in strong contrast to estimates more commonly
adopted in the literature.
The most significant constraints on distance are set by 
resolved radio observations of the previous and current burst
\citep{hjellming1986,taylor1989,rupen2006,obrien2006}.
By assuming that observed radio proper motions (on the sky) can be
ascribed to the fast-moving ejecta or forward shock, workers consistently 
derive a distance of $\sim$1600~pc.   Similarly
large distances were found considering interstellar UV
absorption lines \citep{snijders1987} and HI absorption
measurements \citep{hjellming1986}.  

Given the strength of the evidence, we now consider the implications
of the $d=1600$~pc distance. This distance would rule out the binary
interpretation of the near-IR interferometry data laid out in
\S\ref{binary2}, given existing binary constraints.
Instead, we hypothesize that
the IR emission arises from 
a quasi-stationary\footnote{
It is possible to fit our data with a 
 an expanding wind or jet component, but this 
requires fine-tuning the relative proportions of multiple
components and/or a very asymmetric jetlike emission oriented perpendicular to 
our long (northeast) baselines.  These possibilties will be investigated
more thoroughly in future work with an extended dataset.}
hot gas reservoir that contributes
a combination of emission lines and free-free/bound-free
emission in the NIR bands.  The observed FWHM of $\sim$3 mas is
$\sim$5~AU at 1600~pc, about 3$\times$ the expected RS~Oph binary
separation.  This size is reasonable for a circumbinary disk or
reservoir of hot gas, perhaps kept ionized by the outward moving blastwave
or soft X-ray luminosity from the WD itself following outburst.  This gas
reservoir might be analogous to the ``fallback disk'' inferred to form
after some supernovae \citep[e.g.,][]{wang2006}.

Clearly, the hypothesized gas reservoir must be elongated and somewhat
off-center with respect to the central source in order to fit the
combined IOTA, KI, and PTI interferometry data,
especially the non-zero closure phases.  It is beyond the scope of
this Letter to investigate the details here, and we defer development
of this model to a future paper.

\subsection{Future Work}
While we have ruled out the important class of expanding fireball models
for explaining the IR emission from the recurrent nova
RS~Oph, more work lies ahead to test the other emission mechanisms discussed in
this Letter.  A future study will attempt to synthesize a {\em
  self-consistent} model that can explain the time evolution of the
IR spectrum, NIR and mid-IR interferometer data,
and multi-wavelength light curves at the same time\footnote{Note added: future modeling should address the asymmetric radio nebula and jet observed  
by \citet{obrien2006}.}.

If the close distance $d\simle540$~pc is confirmed, we have a
spectacular opportunity to study in detail a likely SNIa progenitor
and to learn about unexpected shock physics controlling the
non-thermal radio emission.  Alternatively, the further distance
$d\sim1600$~pc suggests we have discovered a significant and new
component to the RS Oph Nova remnant and future work will characterize
the hot circumbinary gas reservoir for the first time.

\acknowledgments {
  We thank J. Sokoloski for productive discussion and
  comments on the manuscript and  members of the AAVSO IR group for 
  early-time photometry.  R.K.B. acknowledges NASA and Dr. Bill Danchi for supporting this work. 
We
  acknowledge SAO, NSF, NASA, CNRS/CNES (France), LAOG, IMEP, and LETI 
  for support of IOTA and IONIC3.
  KI was developed and is
  operated by JPL, MSC and WMKO with funding from NASA.  PTI was
  developed by JPL and is operated by the MSC on behalf of the PTI
  collaboration.  This research made use of SIMBAD, ADS, MSC
  resources, CHARM2, and the 2MASS catalog.  Some data presented
  herein were obtained at the Keck Observatory, operated by a
  scientific partnership among Caltech, UC, and NASA.  }
\bibliographystyle{apj}
\bibliography{apj-jour,RS_Oph,Review2,Review,IONIC3,iKeck,KeckIOTA}


\begin{deluxetable}{cllc}
\tabletypesize{\scriptsize}
\tablecaption{Observing Log for RS~Oph
\label{table_obslog}}
\tablehead{
\colhead{Days since} & \colhead{Date}  & \colhead{Interferometer}  & \colhead{Wavelength} \\
\colhead{2006 Feb 12} & \colhead{(UT)} & \colhead{(Configuration)} &\colhead{($\mu$m)} } 
\startdata
4 & 2006Feb16 & IOTA (A20B15C00)\tablenotemark{a} & 1.65  \\
4 & 2006Feb16 & Keck\tablenotemark{b}            & 2.18  \\
11 & 2006Feb23 & IOTA (A20B15C00) & 1.65  \\
14 & 2006Feb26 & IOTA (A35B15C10) & 1.65 \\
19 & 2006Mar03 & IOTA (A35B15C10) & 1.65 \\
20 & 2006Mar04 & IOTA (A35B15C10) & 1.65 \\
22 & 2006Mar06 & IOTA (A35B15C10)  & 1.65 \\
25 & 2006Mar09 & IOTA (A35B15C10) & 1.65 \\
29 & 2006Mar13 & IOTA (A35B15C10) & 1.65 \\
49 & 2006Apr02 & Palomar Testbed (NW)\tablenotemark{c}  & 2.20 \\
59 & 2006Apr12 & IOTA (A35B15C10) & 1.65 \\
60 & 2006Apr13 & IOTA (A35B15C10) & 1.65 \\
63 & 2006Apr16 & Palomar Testbed (NW) & 2.20 \\
65 & 2006Apr18 & Palomar Testbed (NW) & 2.20 \\
\enddata
\tablecomments{Uniform Disk (UD) diameters of calibrators were generally estimated using {\em getCal}, an SED-fitting routine maintained and distributed by the Michelson Science Center (http://msc.caltech.edu).}
\tablenotetext{a}{IOTA used the following calibrators: HD~152601 (1.6$\pm$0.4 mas), HD~164064 (1.6$\pm$0.5 mas),  $\chi$~UMa \citep[3.24$\pm$0.04 mas;][]{borde2002}, $\rho$~Boo \citep[3.72$\pm$0.12 mas;][]{vanbelle1999}, HD~143033 (1.9$\pm$1.5 mas), HD~156826 (0.6$\pm$0.2 mas), HD~157262 (1.5$\pm$0.5 mas)}
\tablenotetext{b}{KI used the following calibrators:  $\chi$~UMa \citep[3.35$\pm$0.17 mas;][]{cohen1999}, $\rho$~Boo \citep[3.92$\pm$0.19 mas;][]{cohen1999}.}
\tablenotetext{c}{PTI used the following calibrators: HD~164064 (1.6$\pm$0.5 mas), HD~161868 (0.7$\pm$0.1 mas).}
\end{deluxetable}

\begin{deluxetable}{lccc}
\tabletypesize{\scriptsize}
\tablecaption{RS Oph Model Fitting Results
\label{table_model}}
\tablehead{
\colhead{Model} & \colhead{2006 Feb 16 -- 23}  & \colhead{2006 Feb 26 -- Mar 13}  & \colhead{2006 Apr 02 - 18}
     \\
\colhead{Parameter} & \colhead{(Days 4 -- 11)}  &\colhead{(Days 14 -- 29)} & \colhead{(Days 49 -- 65)}}
\startdata
\multicolumn{4}{l}{Gaussian Profile (fitting only to 1.65$\mu$m)} \\
\hline
\qquad FWHM (milliarcseconds) & 3.30 $\pm$ 0.09 & 3.47 $\pm$ 0.03 & 2.87 $\pm$ 0.07 \\
\qquad Reduced $\chi^2$ (V$^2$)       & 0.6             & 1.3             & 1.1 \\
\qquad Reduced $\chi^2$ (CP)       & 1.3             & 3.6             & 5.4 \\
\hline
\multicolumn{4}{l}{Gaussian Profile (fitting only to 2.2$\mu$m)} \\
\hline
\qquad FWHM (milliarcseconds) & 2.56 $\pm$ 0.24 & N/A & 2.00 $\pm$ 0.09 \\
\qquad Reduced $\chi^2$ (V$^2$)       & 0\tablenotemark{a}         & N/A             & 1.0 \\
\qquad Reduced $\chi^2$ (CP)       & N/A             & N/A           & N/A \\
\hline
\multicolumn{4}{l}{Binary Model\tablenotemark{b} (fitting to 1.65$\mu$m \& 2.2$\mu$m data)} \\
\hline
\qquad Separation (milliarcseconds) &                  3.13$\pm$0.12     & 3.23$\pm$0.13 & 3.48$\pm$0.23   \\
\qquad Position Angle\tablenotemark{c} (degs E of N) & 36$\pm$10\tablenotemark{d}        & 45$\pm$5     & 27$\pm$5 \\
\qquad Brightness Ratio\tablenotemark{c} &             0.42$\pm$0.06     & 0.40$\pm$0.06 & 0.21$\pm$0.03 \\
\qquad Reduced $\chi^2$ (V$^2$)  &               0.6                     & 1.7           & 1.2 \\
\qquad Reduced $\chi^2$ (CP)  &              1.1                     & 1.3           & 0.5 \\
\enddata
\tablenotetext{a}{Only one Keck data point available for fitting.}
\tablenotetext{b}{Brightness ratio assumed the same for H and K bands.  Individual components have adopted UD diameters of 0.5~mas.  Here, we considered
only solutions with separations $<$10~mas. }
\tablenotetext{c}{Fainter component with respect to brighter component.}
\tablenotetext{d}{180$\arcdeg$ ambiguity since closure phase data are indistinguishable from zero for this epoch.}
\end{deluxetable}

\end{document}